\documentclass[10pt]{article}

\usepackage{graphics}
\usepackage{graphicx}

\usepackage[a4paper, left=35mm,right=35mm,top=34mm,bottom=34mm]{geometry}
\usepackage[utf8]{inputenc}
\usepackage[T1]{fontenc}
\usepackage[english]{babel}

\usepackage{enumerate}
\usepackage{graphicx}
\usepackage{hyperref}
\hypersetup{
    colorlinks=true,
    linkcolor=blue,
    filecolor=magenta,      
    urlcolor=cyan,
}

\usepackage{mathtools,amsthm,amssymb,amsfonts}
\usepackage{algorithm}
\usepackage{algorithmic}
\makeatother
\theoremstyle{plain}

\theoremstyle{definition}

\theoremstyle{remark}

\usepackage{caption} 
\usepackage[tight,footnotesize]{subfigure}

\usepackage{fancyhdr}
\usepackage{array}
\usepackage{scalefnt}
\usepackage[usenames, dvipsnames]{xcolor}
\usepackage{tikz}
\usepackage{pgfplots}
\usepackage{xcolor}
\usepackage{stmaryrd}

\author{
  {\normalsize Fr\'ed\'eric Magoul\`es}\thanks{CUDA Research Center, \'Ecole Centrale Paris, France
    (correspondence, frederic.magoules@hotmail.com).}
	\and
  {\normalsize Abal-Kassim Cheik Ahamed}\thanks{CUDA Research Center, \'Ecole Centrale Paris, France.}
	\and
  {\normalsize Alban Desmaison, Jean-Christophe L\'echenet, Fran\c cois Mayer, Haifa Ben Salem, Thomas Zhu}\thanks{\'Ecole Centrale Paris, France.}
		}	
\title{Power Consumption Analysis of Parallel Algorithms on GPUs}
\date{}

\begin{document}
\maketitle
\thispagestyle{fancy}

\begin{abstract}
\noindent Due to their highly parallel multi-cores architecture, GPUs are being increasingly used in a wide range of computationally intensive applications. Compared to CPUs, GPUs can achieve higher performances at accelerating the programs' execution in an energy-efficient way. Therefore GPGPU computing is useful for high performance computing applications and in many scientific research fields. In order to bring further performance improvements, GPU clusters are increasingly adopted. The energy consumed by GPUs cannot be neglected. Therefore, an energy-efficient time scheduling of the programs that are going to be executed by the parallel GPUs based on their deadline as well as the assigned priorities could be deployed to face their energetic avidity. For this reason, we present in this paper a model enabling the measure of the power consumption and the time execution of some elementary operations running on a single GPU using a new developed energy measurement protocol. Consequently, using our methodology, energy needs of a program could be predicted, allowing a better task scheduling.
\end{abstract}


\section{Introduction}
\label{sec:introduction}

Many applications such as Computer Vision, e.g., denoising big pictures, Pattern Recognition, High Performance Computing (HPC) or the Monte Carlo methods used in Finance, e.g., exotic path-dependent derivatives pricing using brownian bridge methods, require a high computational density, which strongly depends on the hardware components. The GPU (Graphics Processing Unit) as a highly parallel architecture allows the execution of such complex programs much faster than a CPU (Central Processing Unit)~\cite{creel_high_2012}~\cite{gaikwad_parallel_2010}. Moreover, the development of new graphics parallel programming platforms such as CUDA~\cite{NVidiaCUDACGuide} and OpenCL~\cite{GPU:OpenCL:2010}, has further urged programmers to move to GPUs, in order to take advantage of their high computing capacity~\cite{SKWDKGG2011}.
Most of the top supercomputers in the world include GPUs, as we can observe in the Top 500 list~\cite{Top500}. What makes graphics cards more attractive is their energy consumption/workload ratio. For example, Piz Daint at the Swiss National Supercomputing Centre is the 6th most performing machine in the Top 500 List~\cite{Top500}, however it is ranked 4th in the Green 500 list of November 2013~\cite{Green500}, with 3,185.91 MFLOPS/W power consumption.
The graphics cards remain, however, "\emph{energy-hungry}" compared to the rest of the components of the computer. For example the NVIDIA GTX 780~\cite{GeForce780} reaches a power consumption of 250 W and therefore requires a power supply of 600W.

Since they are based on CMOS transistors, GPUs power consumption essentially consists of static and dynamic power. Static power is increasingly presenting a major concern and is disappeared when the GPU is idle, due to the gate and sub-threshold leakages. Dynamic power, however, is the result of the circuit function and thus depends on the program being executed. In case of non optimal utilization of the GPU cores, performance per watt decreases because of power dissipation on unused or idle cores~\cite{GGGJG2012}.
To reach higher computational density, parallel GPUs in clusters must be optimized and better organized. The energy consumed by a cluster depends both on the architecture of hardware components and the implementation of algorithms of executed programs.
As a consequence, a prediction of the required energy and execution time of an application is a crucial key to optimize the resources allocation of a cluster, and thereafter find the most adequate architecture and its corresponding tasks scheduling~\cite{LDHDBX2011}~\cite{HSHW2012}.

In this work we propose an evaluation methodology of the operating mode of the GPU during the execution of some computing applications focusing on its energy consumption behavior, which enables to better understand the energetic specifications of GPUs. For this purpose, we built an experimental device in order to measure the power consumed by the graphics card uniquely.

In order to validate and consolidate the proposed protocol, we propose in this paper to analyze and evaluate the energy efficiency of elementary operations such as addition, product, etc. After analyzing the measures we obtained, we propose a mathematical model able to determine the energy consumption as well as the time execution of GPU programs. After that, we can study the variation of the power dissipation and time execution factors depending on the chosen input parameters, in order to avoid the underutilization of the GPU resources.
We have tested our algorithms with real data arising from the gravity equations, as described in~\cite{cheikahamed:2013:inproceedings-3} an \cite{ahamed2013stochastic}.

This paper is organized as follows. The plan is divided into two main parts.
In Section~\ref{sec:measuring_power_consumption_of_GPUs} we explain how the energy consumption is measured, and describe the necessary hardware components required to design the experiment.
We also introduce the experimental protocol that we have designed to acquire accurate measurements of the disappeared power from a GPU.
Section~\ref{sec:results_and_Prediction_algorithm} presents both numerical results and the model we propose.

\section{Measuring Power Consumption of GPUs}
\label{sec:measuring_power_consumption_of_GPUs}

\subsection{Context}
\label{subsec:context}

Commercialized since 1999\footnote{world's first GPU, the GeForce 256, was released on October 1999}, \emph{Graphics Processing Units} were designed to perform always more 3D graphics computations in order to offload massive parallelizable computations from \emph{Central Processing Units}. As a consequence, the growing rate of the computing power of GPUs was directly related to how graphically advanced video games were at the time. On the other hand, cores of CPUs were slowly reaching their clock rate limit\footnote{around $4$ gHz as usually recorded in B2C market}~\cite{SUTTER2005}. Contrary to CPUs before 2001\footnote{IBM commercialized Power4, the very first dual core CPU, in 2001}, GPUs have always had a multi-cores architecture, which help them manage the thermal dissipation coming from the Joule effect. Thus, the gap (in terms of Floating Operations Per Second) between the computing power of GPUs and that of CPUs was increased. With time, even though no API for GPU yet existed, the concept of \emph{GPGPU}\footnote{General Purpose Graphics Processing Unit} emerged as the possibility of making general computations on GPUs became increasingly interesting.

In 15 February 2007, the first nVidia GPGPU toolkit named \emph{CUDA} (\emph{Compute Unified Device Architecture}) toolkit was publicly released. Based on C/C++ programming language, the Application Programming Interface (\emph{API})  provided a standard on how to write code to be launched from CPU/GPU and executed on a GPU. Most source codes using \emph{CUDA} are divided in multiple steps as follows: (i) allocation of memory on the CPU-side, (ii) allocation of memory on the GPU-side, (iii) copying allocated values from CPU to GPU, (iv) kernel launch\footnote{code launched from CPU and executed on GPU}, (v) Copy results from GPU to CPU, (vi) deallocation of allocated memory.
where the copy operations use the motherboard bus to transfer data between the \emph{Central Memory} (\emph{CM}) of the computer and the \emph{Global Memory} inside the GPU. Programming on GPU requires a careful manual memory management. Furthermore, writing an efficient code requires the use of many types of memory such as \emph{shared memory} ("shared" among streaming multiprocessors), \emph{registers} (local to each core) or \emph{global memory} (shared by all streaming multiprocessors). Consequently, until the release of \emph{unified memory}~\cite{UnifiedMemory} in CUDA 6.0 (on April 15 2014), managing memories was one of the hardest parts of CUDA programming.

GPUs are interconnected to a motherboard using a standardized type of slot. Before 2004, three types of GPU slots were used: \emph{AGP}, \emph{PCI} and \emph{PCI-X}. The need for a higher system bus throughput at an acceptable price leads to the creation of a new serialized slot: the \emph{PCI express}~\cite{PCIe3.0, PCIeArchi} (\emph{PCIe}). In the case of GPGPU, both the GPU's expansion slot system bus throughput and the motherboard bus speed had a direct impact on the time spent in steps (iii) and (v), i.e., data transfers between CPU-side and GPU-side.
Compared to the highest system bus throughput attained before 2004 (AGP maximum throughput goes up to $2133$ MB/s), PCIe x16 is approximately $15$ times faster for the future v4.0~\cite{PCIe4.0} version of the standard.
Some PCIe lanes are especially dedicated for power supply. More particularly, GPUs usually drain power from the PCIe bus and, for high consuming ones, directly from the \emph{Power Supply Unit} (\emph{PSU}).

Although GPUs are very energy-greedy (up to several hundreds of Watts), comparing the power consumption of different parallelized algorithms on GPUs is seldom done.
A reason for that is the absence of on-chip captors to get the real time power consumption of GPUs. The first step therefore was to find a way to measure that consumption.
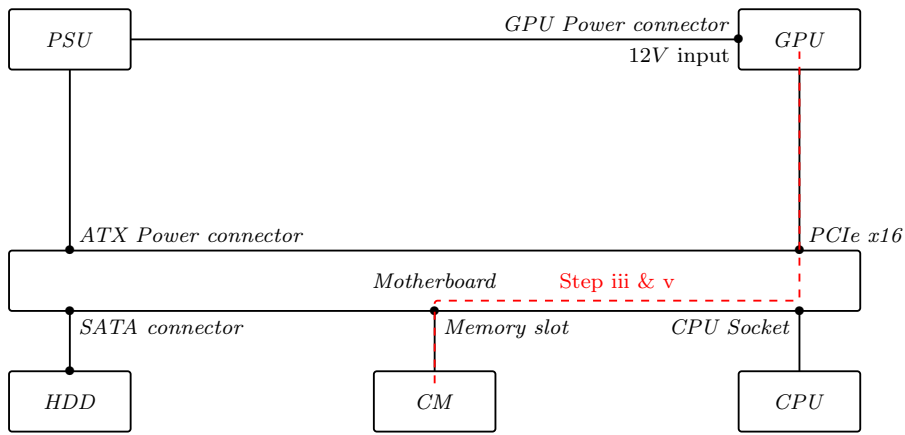
\begin{figure}[!ht]
\centering
{\scalefont{0.8}
\begin{tikzpicture}[scale = 0.80, line width = 0.7pt]
\draw[rounded corners=0.4mm] (0,0) rectangle (2,1);
\draw (1,0.5) node{\emph{HDD}};
\draw[rounded corners=0.4mm] (6,0) rectangle (8,1);
\draw (7,0.5) node{\emph{CM}};
\draw[rounded corners=0.4mm] (12,0) rectangle (14,1);
\draw (13,0.5) node{\emph{CPU}};
\draw[rounded corners=0.4mm] (0,2) rectangle (14,3);
\draw (7,2.5) node{\emph{Motherboard}};
\draw[rounded corners=0.4mm] (0,6) rectangle (2,7);
\draw (1,6.5) node{\emph{PSU}};
\draw[rounded corners=0.4mm] (12,6) rectangle (14,7);
\draw (13,6.5) node{\emph{GPU}};
\draw (1,1) -- (1,2);
\draw (7,1) -- (7,2);
\draw (13,1) -- (13,2);
\draw (1,3) -- (1,6);
\draw (13,3) -- (13,6);
\draw (2,6.5) -- (12,6.5);
\draw (1,1) node{$\bullet$};
\draw (1,2) node{$\bullet$} node[below right]{\emph{SATA connector}};
\draw (1,3) node{$\bullet$} node[above right]{\emph{ATX Power connector}};
\draw (7,2) node{$\bullet$} node[below right]{\emph{Memory slot}};
\draw (13,2) node{$\bullet$} node[below left]{\emph{CPU Socket}};
\draw (13,3) node{$\bullet$} node[above right]{\emph{PCIe x16}};
\draw (12,6.5) node{$\bullet$} node[above left]{\emph{GPU Power connector}} node[below left]{$12 V$ input};
\draw[dashed, red, rounded corners=0.4mm] (7,0.8) -- (7,2.17) -- (13,2.17)node[midway, above]{Step iii \& v} -- (13,6.3);
\end{tikzpicture}}
\caption{Data transfers between CPU-side and GPU-side}
\end{figure}

\subsection{Experimental Protocol}
\label{subsec:experimental_protocol}

\subsubsection{Building a Physical Device}

All the measuring devices we encountered were not satisfying as they took into account the consumption of the whole computer. To achieve a good precision, we needed to measure the consumption of the GPU alone and we therefore decided to build our own measuring device.
The GPU we used had two physical connections to the computer. It received energy from a power unit, and was also directly connected to the motherboard. Since we obviously could not access the flow from within the circuit to measure it, we decided to use amperometric clamps. An amperometric clamp is a device which measures the intensity of a current by using the magnetic field it creates. The clamp then delivers a tension which can be measured using an oscilloscope. From that tension, it is easy to go back to the intensity as both are linearly related.

Accessing the power transmitted to the GPU via the power unit therefore became straightforward. Using an amperometric clamp to measure the intensity $ I_{PU} $ of the current coming from the power unit, and knowing that the power unit delivered a constant voltage of $12V$, the power was $P_{PU}  = 12 * I_{PU}$.
Measuring the second component of the power, the one coming directly from the motherboard was however much more difficult: there were no cables between the GPU and the motherboard, and therefore no place to clip the clamp. We therefore had to use a \emph{PCI Express Riser 16x} to connect the GPU to the motherboard, using it as an extension cable.
\begin{figure}[!ht]
\centering
\includegraphics[scale=0.55]{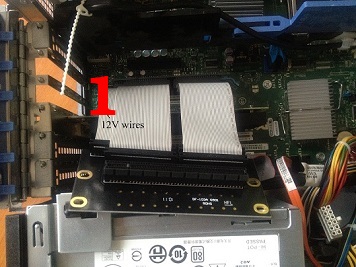}
\caption{A PCI Express Riser}
\label{fig:img:riser_wires}
\end{figure}
The riser has three connector pins through which the 12V voltage flows, and three connector pins through which the 3.3V voltage flows~\cite{PCI_Express}. Figure~\ref{fig:img:riser_wires} illustrates this by showing the 12V wires (1). What's more, according to the specifications of the riser, the voltages delivered are constant so measuring the intensity of the current immediately gives us the power.

\begin{figure}[!ht]
\centering
\includegraphics[scale=0.45]{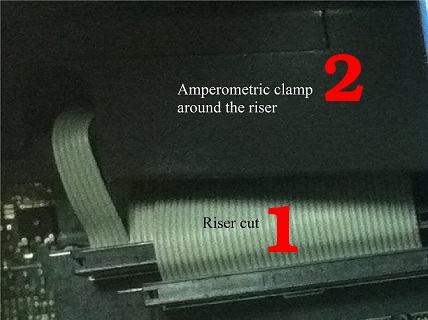}
\caption{Clamp clipped around the cut riser}
\label{fig:img:riser_cut}
\end{figure}

\begin{figure}[!ht]
\centering
{\scalefont{0.8}
\begin{tikzpicture}[scale = 0.80, line width = 0.7pt]
\draw[rounded corners=0.4mm] (0,0) rectangle (2,1);
\draw (1,0.5) node{\emph{HDD}};
\draw[rounded corners=0.4mm] (6,0) rectangle (8,1);
\draw (7,0.5) node{\emph{CM}};
\draw[rounded corners=0.4mm] (12,0) rectangle (14,1);
\draw (13,0.5) node{\emph{CPU}};
\draw[rounded corners=0.4mm] (0,2) rectangle (14,3);
\draw (7,2.5) node{\emph{Motherboard}};
\draw[rounded corners=0.4mm] (0,6) rectangle (5,7);
\draw (2.5,6.5) node{\emph{Power Supply Unit}};
\draw[rounded corners=0.4mm] (12,6) rectangle (14,7);
\draw (13,6.5) node{\emph{GPU}};
\draw (1,1) -- (1,2);
\draw (7,1) -- (7,2);
\draw (13,1) -- (13,2);
\draw (1,3) -- (1,6);
\draw[blue, line width = 0.9pt] (13,3) -- (13,4.5)node[near end]{\emph{PCI x16 Riser}};
\draw[blue, line width = 0.9pt] (13,4.7) -- (13,5.5);
\draw (13,5.5) -- (13,6);
\draw (13,5.5) node{$\bullet$} node[left]{\emph{PCIe x16 in}};
\draw (5,6.5) -- (6.8,6.5);
\draw (7,6.5) -- (12,6.5);
\draw (12,6.5) node[above left]{$12 V$ \emph{input}};
\draw (1,1) node{$\bullet$};
\draw (1,2) node{$\bullet$};
\draw (1,3) node{$\bullet$};
\draw (7,2) node{$\bullet$};
\draw (13,2) node{$\bullet$};
\draw (13,3) node{$\bullet$} node[above left]{\emph{PCIe x16}};
\draw (12,6.5) node{$\bullet$};
\draw[rounded corners=0.4mm, red] (3,4) rectangle (5,5);
\draw (4,4.5) node{\emph{LAN}};
\draw[rounded corners=0.4mm, red] (6,4) rectangle (11,5);
\draw (8.5,4.5) node{\emph{Oscilloscope}};
\draw[red] (5,4.5)node{\textcolor{red}{$\bullet$}} -- (6,4.5)node{\textcolor{red}{$\bullet$}};
\draw[PineGreen] (11,4.5)node{\textcolor{red}{$\bullet$}} -- (12.8,4.5);
\draw[PineGreen] (12.8,4.5) arc (180:-60:0.2 and 0.15);
\draw[PineGreen] (12.8,4.5) arc (180:240:0.2 and 0.15);
\draw[PineGreen] (7,5)node{\textcolor{red}{$\bullet$}} -- (7,6.3)node[midway, left]{\emph{Amperometric Clamp}};
\draw[PineGreen] (7,6.3) arc (270:30:0.15 and 0.2);
\draw[PineGreen] (7,6.3) arc (270:330:0.15 and 0.2);
\draw[red] (4,3)node{\textcolor{red}{$\bullet$}} -- (4,4)node{\textcolor{red}{$\bullet$}}node[midway, left]{\emph{Ethernet Cable}};
\end{tikzpicture}}
\caption{Experimental Protocol}
\label{EP}
\end{figure}
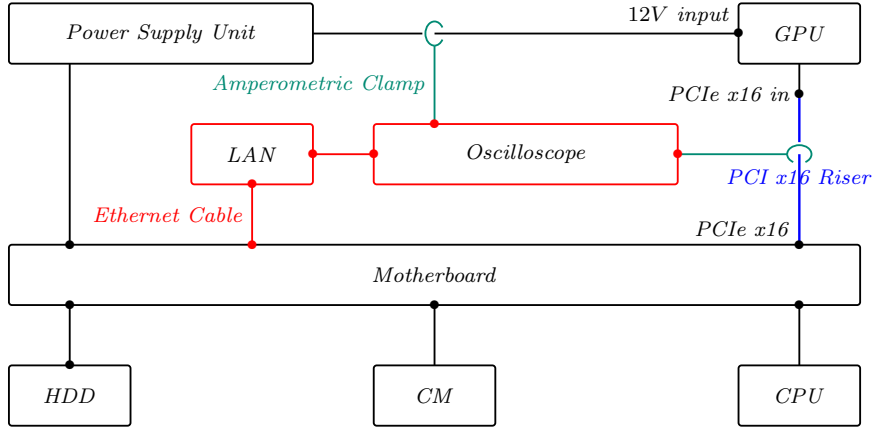

We therefore used two amperometric clamps, one to measure the intensity of the current flowing into each set of wire. However, as can see in Figure~\ref{fig:img:riser_wires}, there was still no way to clip the clamps on the wires so we had to cut them, as described in Figure~\ref{fig:img:riser_cut} where (1) correponds to the riser cut and (2) the amperometric clamp around the riser. 
The power given by the motherboard to the GPU was then given by the formula $P = 3.3 \times I_{3.3} + 12 \times I_{12}$.
The measures showed that the variation of the intensity of the current going through the 3.3V wires was very small compared to that going through the 12V wires, so we neglected that variation and considered that the power brought through the 3.3V wires was constant and equal to 1W. This enabled us to use one less amperometric clamp, which also simplified the physical setting up of the device. Figure~\ref{EP} gives a summary of the physical implementation of the device.

\subsubsection{Sample Code Used}

We then wanted to measure the power consumption of the basic operations a GPU can perform, which includes memory allocation, sum, product, and communication. To do so, we created a small program which did the following. The program steps are: (i) starts, inserting the size of the vectors it is going to work on, (ii) randomly generates a vector of that specific size using the CPU and stores it on the RAM, (iii) asks the user for the parameters he wants to use: number of sums, number of products, number of threads per block and block load factor, (iv) allocates memory on the GPU RAM, (v) pauses so as to measure the consumption of only allocating memory on the GPU without using it. To pause, the program asks the user the number of copies he wants to do, (vi) copies the same vector on the GPU RAM the correct number of times, (vii) does the right number of sums and prints the time elapsed, (viii) does the right number of products and prints the time elapsed, (ix) copies all these results back to the CPU and returns them.

\section{Numerical Results and Prediction Algorithm}
\label{sec:results_and_Prediction_algorithm}

References \cite{cheikahamed:2012:inproceedings-2} and \cite{cheikahamed:2012:inproceedings-1} clearly proved that the implmentation of linear algebra opeations is more efficient on GPU that on CPU.
However, given that these operations are a set of linear combination of elementary operations (addition, multiplication, ...), we test our protocol on them.

\subsection{Validation of the experimental protocol}

Before we can apply our protocol and collect measures, a simple verification that our method is correct is to check that different types of operations on the GPU give significantly different voltage levels. Indeed, if whatever the calculation made on the GPU, the consumption is the same or, more generally, if there is no correlation between the operations made on the GPU and the consumption observed on the oscilloscope, our endeavor to build our model is pointless.
\begin{figure}[!ht]
\centering
\includegraphics[scale=0.55]{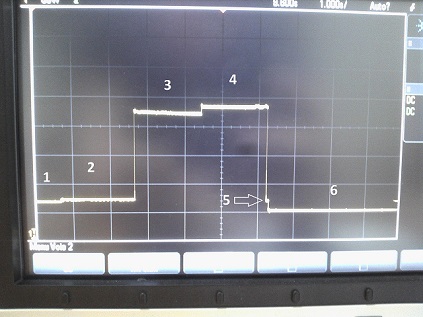}
\caption{A curve seen on the oscilloscope when our program runs on the GPU}
\label{fig:img:oscilloscope_example}
\end{figure}
Figure~\ref{fig:img:oscilloscope_example} shows an example of the curves observed on the oscilloscope when the program described above is executed on the GPU. In this figure, we can clearly distinguish different parts that can be associated with a certain type of operation, which confirms our method. More precisely, we can see six levels :
(1) allocation of the memory on the GPU, no copy made; (2) copies from the CPU to the GPU, (3) sums on the GPU, (4) products on the GPU, (5) copies from the GPU to the CPU, and deallocation of the memory on the GPU, (6) stand-by.
After a few moments, if the GPU stays idle, it can reach an energy saving mode. In that case, the consumption decreases yet again.

\subsection{Collection of the measures}

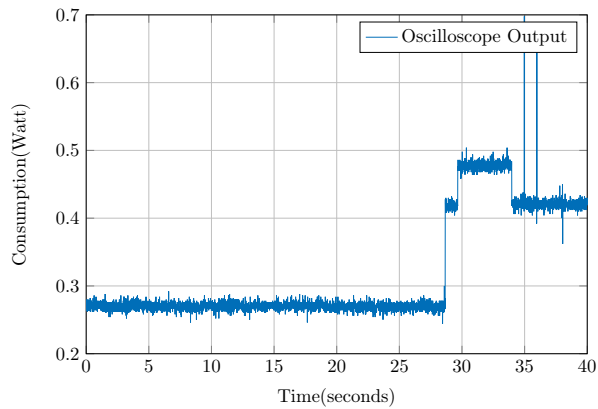
\begin{figure}[!ht]
  \begin{center}
    \begin{tikzpicture}[scale=0.7]
      \begin{axis}[
        height=8cm,
        width=11cm,
        xmin = 0, xmax = 40,
        ymin = 0.2, ymax = 0.7,
        xlabel=Time(seconds), ylabel=Consumption(Watt),
        xmajorgrids, ymajorgrids, scaled x ticks=false
      ]
      \addplot[NavyBlue, line width=0.2pt] file {data/Oscilloscope.txt};
      \addlegendentry{Oscilloscope Output}
      \end{axis}
    \end{tikzpicture}
  \end{center}
  \caption{An Exported Curve}
  \label{oscilloscope_curve}
\end{figure}

By executing the same program with different parameters, we can get multiple curves on the oscilloscope. To manipulate the measures, we need to send them to a computer. With our oscilloscope, we could export the curves displayed on the screen in the form of csv files. Figure~\ref{oscilloscope_curve} is an example of a curve reconstructed from such a csv file.

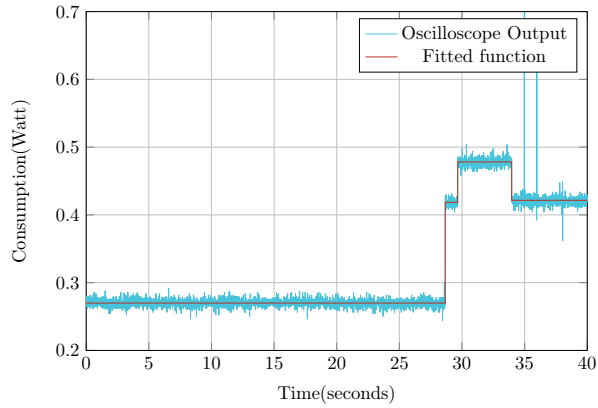
\begin{figure}[!ht]
  \begin{center}
    \begin{tikzpicture}[scale=0.7]
      \begin{axis}[
        height=8cm,
        width=11cm,
        xmin = 0, xmax = 40,
        ymin = 0.2, ymax = 0.7,
        xlabel=Time(seconds), ylabel=Consumption(Watt),
        xmajorgrids, ymajorgrids, scaled x ticks=false
      ]
      \addplot[SkyBlue, line width=0.2pt] file {data/Oscilloscope.txt};
      \addlegendentry{Oscilloscope Output}
      \addplot[BrickRed, line width=0.5pt] file {data/StepApprox.txt};
      \addlegendentry{Fitted function}
      \end{axis}
    \end{tikzpicture}
  \end{center}
  \caption{The step function approximating the curve}
  \label{step_curve}
\end{figure}

To analyze more easily these curves, which are approximations of step functions, we wrote a script to automatically detect their different phases and compute the corresponding step function.
The result of this process on the curve in Figure~\ref{oscilloscope_curve} is shown in Figure~\ref{step_curve}.

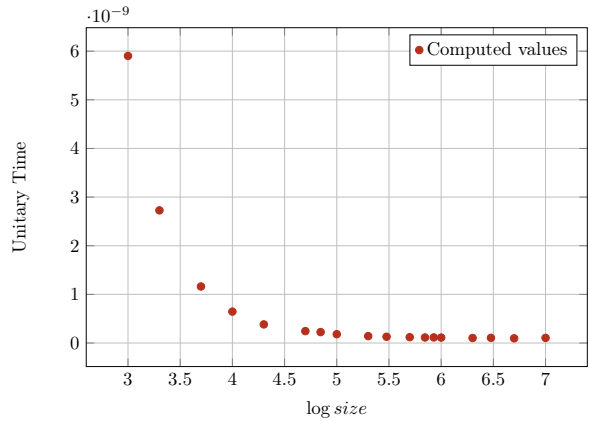
\begin{figure}[!ht]
  \begin{center}
    \begin{tikzpicture}[scale=0.7]
      \begin{axis}[
        height=8cm,
        width=11cm,
        xlabel=$\log size$, ylabel=Unitary Time,
        xmajorgrids, ymajorgrids, scaled x ticks=false
      ]
      \addplot[BrickRed, only marks, mark = *] file {data/SumRegression.txt};
      \addlegendentry{Computed values}
      \end{axis}
    \end{tikzpicture}
  \end{center}
  \caption{Unitary Time (in seconds) for a sum}
  \label{UnitTime}
\end{figure}

With that treatment, we can measure for each execution of our program the duration and the power consumption of each type of operation.
We used our measures to plot the time needed to do one addition against the size of the vector used. The graph can be seen in Figure~\ref{UnitTime}.

This graph represents the time spent on each addition (time to compute the result divided by the size of the input vector) given an input vector size.
As expected, the bigger the input vector, the more efficient the GPU is. This is true until we reach the point where all the streaming multiprocessors are used. The execution time is then proportional to the input vector size, which explains the horizontal line on the graph.

\begin{figure}[!ht]
  \begin{center}
    \begin{tikzpicture}[scale=0.7]
      \begin{axis}[
        height=8cm,
        width=11cm,
        xlabel=$\log size$, ylabel= Unitary Consumption(Watt),
        xmajorgrids, ymajorgrids, scaled x ticks=false
      ]
      \addplot[BrickRed, only marks, mark = *] file {data/UnitaryPower.txt};
      \addlegendentry{Computed values}
      \end{axis}
    \end{tikzpicture}
  \end{center}
  \caption{Unitary Consumption (in Watt) for a sum}
  \label{UnitPower}
\end{figure}
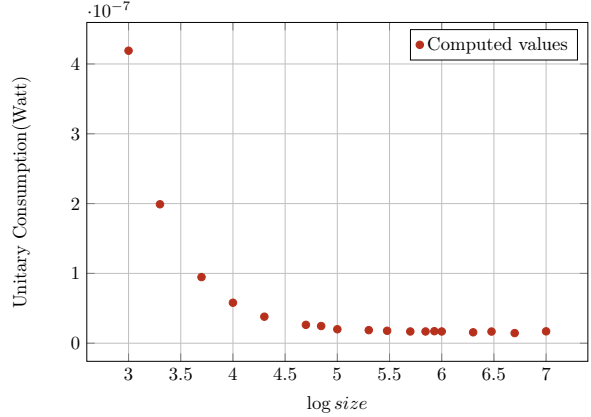

Another series of graph we were able to plot showed the power spent to perform one elemental action. Figure~\ref{UnitPower} illustrates that for the sum.

We find that from an energetic point of view too, using bigger vectors is far more efficient than using small ones and as for the time, the unitary power decreases until the vectors as big enough to use all the streaming multiprocessors.

\subsection{Regression: Generalized Least Squares (GLS) Method}

We then did regressions on our data to obtain a model which can be used to predict the time and energy consumption of an algorithm knowing its number of elementary operations.

Given $(x,y)={(x_i,y_i)}_{i\in\llbracket 1,n\rrbracket}$ we are looking for a function $f$ such as for all $i\in\llbracket 1,n\rrbracket$: $y_i = f(x_i) + \epsilon_i$ where $\epsilon = (\epsilon_i)_{i\in\llbracket 1,n\rrbracket}$ is a sequence of centered, independant, identically distributed variables with variance $\sigma$. Furthermore $\epsilon$ corresponds to the intrinsic errors of measure coming from the sensors used,
So we can\footnote{This is not always right for quantum physics} suppose that $\epsilon$ is independant of $x$. In our case, $x$ corresponds to the logarithm of the size of a vector while $y$ depends on the measures (unitary time or unitary consumption). Looking at the shape of Figures~\ref{UnitTime} and~\ref{UnitPower}, we see that we are actually looking for a decreasing kind of regression function. Hence, we used a linear combination of the most usual decreasing functions, i.e., $x\mapsto 1/x^k$:
\[
f:x\mapsto \sum_{k=1}^{N}\frac{a_k}{x^k} + a_\infty
\]
Lets denote $X$ the matrix such as (i.e. "featuring" step of our regression):
\[
X = \begin{bmatrix}
1/x_1^1 & \cdots & 1/x_1^N & 1 \\
\vdots & \cdots & \vdots & \vdots \\
1/x_n^1 & \cdots & 1/x_n^N & 1
\end{bmatrix}
\]
Thus, the regression parameters, for a given $N$ (which corresponds to a fitting degree of complexity) are $a = (a_1,\cdots,a_n,a_\infty)$. The general penalized regression~\cite{STEPH2014} problem is of the form\footnote{We equivalently have $\hat{f}_N:x\mapsto \sum_{k=1}^{N}\frac{\hat{a}_k}{x^k} + \hat{a}_\infty$}
\[
\hat{a} = \underset{a}{argmin}\left(\vert\vert y-X*a\vert\vert_1 + \lambda Pen(a) \right)
\]
Furthermore, $\vert\vert y-X*a\vert\vert_1$ corresponds to a Goodness-Of-Fit while $\vert\vert .\vert\vert_2$ is the penalization function. The $\vert\vert .\vert\vert_1$ is freely choosen depending on the regression type while $Pen$ depends on what kind of result we are expecting, e.g., taking $Pen = a\mapsto\sum_{k=1}^{N}\vert a_k\vert + \vert a_\infty\vert$ will induce sparsity of $\hat{a}$) and $\lambda$ the weight of the penalization in the regression.

For the sake of simplicity, we chose $\lambda = 0$ and $\vert\vert .\vert\vert_1$ the standard Euclidian norm. This corresponds to the Generalized Least Squares Method where explicit solutions exist\footnote{We will sometimes write $\hat{a}$ instead of $\hat{a}^{(n)}$ in order to make it easier to read}
\[
\hat{a} = (X'X)^{-1}X'y
\]
More particularly, $X\hat{a}$ is the Euclidian projection of $y$ in the subspace generated by the columns of $X$. This gives us the $L^2(\Omega,\mathcal{F},\mathbb{P})$-Pythagoras relation\footnote{"Total dispersion" $=$ "The dispersion around the regression" + "The dispersion caused by the regression"}
\[
\vert\vert y-\bar{y}\vert\vert_1^2 = \vert\vert y-X\hat{a}\vert\vert_1^2 + \vert\vert \hat{a}-\bar{y}\vert\vert_1^2
\]
Then the proportion of variance explained is
\[
R^2 = \vert\vert y-X\hat{a}\vert\vert_1^2/\vert\vert y-\bar{y}\vert\vert_1^2
\]
The $R^2$ will be our regression "precision" measure (the higher the $R^2$ is the better). About the convergence of the estimator of $a$, the theory of GLS says that $\hat{a}$ is an unbiased, consistent, efficient, asymptotically normal estimator and that:
\[
\sqrt{n}(\hat{a}^{(n)}-a) \underset{n\rightarrow +\infty}{\overset{\mathcal{L}}{\longrightarrow}} \mathcal{N}_{N+1}(0,(X'\Sigma X)^{-1})
\]
with $\Sigma := (Cov(\epsilon_i,\epsilon_j))_{1\leq i,j\leq n} = \sigma I_n$. As $(X' X)^{-1}$ is symetric definite positive, there exist (Cholesky decomposition) a lower triangular matrix $L$ and a diagonal matrix $D$ (without zeros in his diagonal) such as $(X' X)^{-1} = LDL' = L\sqrt{D}(L\sqrt{D})'$ with $\sqrt{D}$ the matrix with squared diagonal (element by element). Then $\mathcal{N}_{N+1}(0,(X'\Sigma X)^{-1}) \overset{\mathcal{L}}{=} L\sqrt{D/\sigma}\mathcal{N}_{N+1}(0,I_{N+1})$. Thus, by denoting $B_{N+1}^\infty(0,1)$ the open unit $N\!+\!1\!-\!ball$, i.e., open unit ball in dimension $N+1$, associated to the $\sup$ norm and taking the following opened convex neighbourhood\footnote{as a the image of the open unit ball by the homeomorphism $u\mapsto \hat{a}^{(n)} + L\sqrt{D/(n\hat{\sigma}^{(n)})}u$} of $\hat{a}^{(n)}$
\[
\tilde{D}_n(0,r_\alpha) := \hat{a}^{(n)}+L\sqrt{D/(n\hat{\sigma}^{(n)})}B_{N+1}^\infty\left(0,r_\alpha\right)
\]
we deduce that $\tilde{D}_n(0,r_\alpha)$ is an asymptotic confidence domain of level $1-\alpha = (2\phi(r_\alpha)-1)^{N+1}$ (where $\phi$ is the cumulated distribution of $\mathcal{N}(0, 1)$  with $\hat{\sigma}^{(n)}$ an empirical estimator of the variance of $\epsilon_1$. Thus,
\[
\tilde{D}(0, \underset{r_\alpha}{\underbrace{q_{\mathcal{N}(0,1)}^{((1-\alpha)^{1/(N+1)}+1)/2}}})
\]
is a confidence domain for $a$ of level $1-\alpha$. However, the computation of ($L$ and $D$) the confidence domain is left over for readers.

On the other hand, for a given $z := \log(size_z)\in\mathbb{R}$, we will be able to predict the time spent on a given operation, i.e., it will be $f_N(z) = f_N(\log(size_z))$).

\subsubsection{Unitary Time Regression}

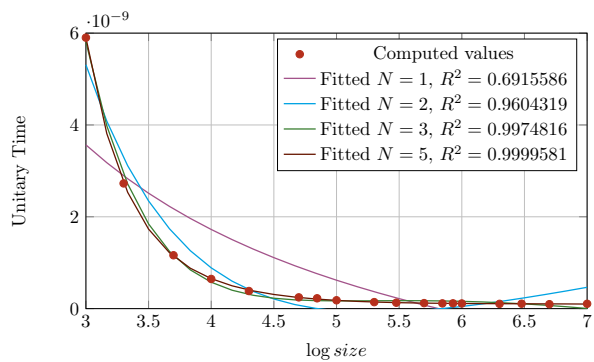
\begin{figure}[!ht]
  \begin{center}
    \begin{tikzpicture}[scale=0.7]
      \begin{axis}[
        height=6.8cm,
        width=11.00263cm,
        xmin = 3, xmax = 7,
        ymin = 0, ymax = 6E-09,
        xlabel=$\log size$, ylabel=Unitary Time,
        xmajorgrids, ymajorgrids, scaled x ticks=false
      ]
      \addplot[BrickRed, only marks, mark = *] file {data/SumRegression.txt};
      \addlegendentry{Computed values}
      \addplot[DarkOrchid, line width = 0.7pt, domain=3:7] {2.2099914856800486E-8 * x^-1 + -3.8009730591705355E-9};
      \addlegendentry{Fitted $N = 1$, $R^2 = 0.6915586$}
      \addplot[Cerulean, line width = 0.7pt, domain=3:7] {-9.735453234983343E-8 * x^-1 + 2.5788445217705575E-7 * x^-2 + 9.108921796113016E-9};
      \addlegendentry{Fitted $N = 2$, $R^2 = 0.9604319$}
      \addplot[OliveGreen, line width = 0.7pt, domain=3:7] {2.02832389645393E-7 * x^-1 + -1.0752502679701515E-6 * x^-2 + 1.896143146635804E-6 * x^-3 + -1.255705816272474E-8};
      \addlegendentry{Fitted $N = 3$, $R^2 = 0.9974816$}
      \addplot[Sepia, line width = 0.7pt, domain=3:7] {2.552185568785448E-7 * x^-1 -2.6519199128882676E-6 * x^-2 + 1.377818601033809E-5 * x^-3 -3.620192787590071E-5 * x^-4 + 3.931483477481734E-5 * x^-5 -9.670487968130997E-9};
      \addlegendentry{Fitted $N = 5$, $R^2 = 0.9999581$}
      \end{axis}
    \end{tikzpicture}
  \end{center}
  \caption{Time in seconds for a sum: Regression}
  \label{UnitaryTimeRegression}
\end{figure}

Figure~\ref{UnitaryTimeRegression} collects the different regressions of unitary time obtained for different fitting degrees of complexity.
As we can see in Figure~\ref{UnitaryTimeRegression}, the regression accuracy is suitable for $N=5$. As a result, the parameters of the selected regression, i.e., $N=5$, are given in the following.
\[
  \left\{
  \begin{array}{cccccccc}
  &\hat{a}_1 &= &0.0000003 &\hat{a}_2 &= &-0.0000027\\
  &\hat{a}_3 &= &0.0000138 &\hat{a}_4 &= &-0.0000362\\
  &\hat{a}_5 &= &0.0000393 &\hat{a}_\infty &= &-9.67\times 10^{-9}\\
  \end{array}
  \right.
\]

\subsubsection{Unitary Power Regression}

\begin{figure}[!ht]
  \begin{center}
    \begin{tikzpicture}[scale=0.7]
      \begin{axis}[
        height=6.8cm,
        width=11.00263cm,
        xmin = 3, xmax = 7,
        ymin = 0, ymax = 4.5E-07,
        xlabel=$\log size$, ylabel=Unitary Consumption,
        xmajorgrids, ymajorgrids, scaled x ticks=false
      ]
      \addplot[BrickRed, only marks, mark = *] file {data/UnitaryPower.txt};
      \addlegendentry{Computed values}
      \addplot[DarkOrchid, line width = 0.7pt, domain=3:7] {1.5494595400323655E-6 * x^-1 + -2.5743414725331627E-7};
      \addlegendentry{Fitted $N = 1$, $R^2 = 0.7017766$}
      \addplot[Cerulean, line width = 0.7pt, domain=3:7] {-6.657123621939238E-6 * x^-1 + 1.7716797092615685E-5 * x^-2 + 6.294823989253881E-7};
      \addlegendentry{Fitted $N = 2$, $R^2 = 0.9637507$}
      \addplot[OliveGreen, line width = 0.7pt, domain=3:7] {1.3140123132090488E-5 * x^-1 + -7.020307908219697E-5 * x^-2 + 1.2505013045008978E-4 * x^-3 + -7.993831511196679E-7};
      \addlegendentry{Fitted $N = 3$, $R^2 = 0.9970168$}
      \addplot[Sepia, line width = 0.7pt, domain=3:7] {4.527807344256729E-5 * x^-1 -4.420569095699989E-4 * x^-2 + 0.00213149480634911 * x^-3 -0.0051157195265432165 * x^-4 + 0.004973491938471852 * x^-5 -1.8106915811166857E-6};
      \addlegendentry{Fitted $N = 5$, $R^2 = 0.9998981$}
      \end{axis}
    \end{tikzpicture}
  \end{center}
  \caption{Power consumption (in Watt) for a sum: Regression}
  \label{UnitaryPowerRegression}
\end{figure}
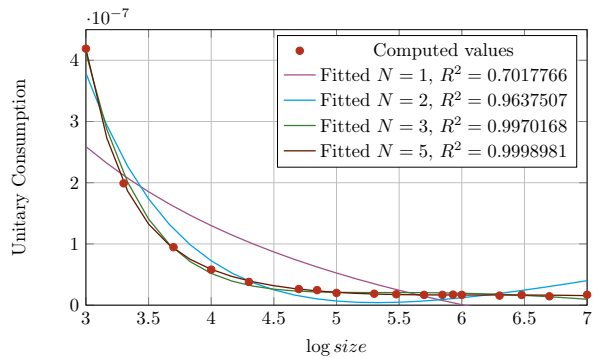

The different regressions of unitary power for different fitting degrees of complexity are drawn in Figure~\ref{UnitaryPowerRegression}.
Figure~\ref{UnitaryPowerRegression} shows that the regression precision is correct for $N=5$.
The coefficients of the selected regression, i.e., $N=5$, are described in the following.
\[
  \left\{
  \begin{array}{ccccccc}
  &\hat{a}_1 &= &0.0000453 &\hat{a}_2 &= &-0.0004421\\
  &\hat{a}_3 &= &0.0021315 &\hat{a}_4 &= &-0.0051157\\
  &\hat{a}_5 &= &0.0049735 &\hat{a}_\infty &= &-0.0000018\\
  \end{array}
  \right.
\]

\subsection{Model and Prediction}

One of our main objectives was to be able to predict, with some approximations, the total energy consumption of a program running on a GPU.

To achieve this goal, we first need to split the initial program in a set of elemental operations. This decomposition enables us to get the total number of each operation the program will order the GPU to do.
The basic inputs that we considered in our very simplified model were the size of the data that needs to be copied to the GPU, the size of memory used on the GPU, and the number of total additions and multiplications performed by the program.
With all these data, we tried to predict the time needed for the program to execute and the total energy consumed during the program execution.

To get this model, we used few simple equations that allowed us to get the time, the energy consumption of the calculation and the energy consumption of other status of the GPU:
\begin{eqnarray*}
T_{total} &=& T_{CPU} \\
      &+& {\hat{f}_N^{time,copy}}(\log(n)) * n \\
      &+& {\hat{f}_N^{time,sum}}(\log(n)) * n \\
      &+& {\hat{f}_N^{time,product}}(\log(n)) * n
\end{eqnarray*}
where $n$ is the size of the vectors, ${\hat{f}_N^{time,copy}}$ the unitary transfer computing time function, ${\hat{f}_N^{time,sum}}$ the unitary running time of the addition of vectors function and ${\hat{f}_N^{time,product}}$ the unitary time of the term-to-term product function, like a dot product without the addition of all the results.

\begin{eqnarray*}
E_{total} &=& P_{idle} * T_{idle} \\
          &+& P_{active} * T_{active} \\
          &+& P_{pause} * T_{pause} \\
          &+& P_{power} * T_{power} \\
          &+& {\hat{f}_N^{power,copy}}(\log(n))* {\hat{f}_N^{time,copy}}(\log(n))* n^2 \\
          &+& {\hat{f}_N^{power,sum}}(\log(n)) * {\hat{f}_N^{time,sum}}(\log(n)) * n^2 \\
          &+& {\hat{f}_N^{power,product}}(\log(n)) * {\hat{f}_N^{time,product}}(\log(n)) * n^2
\end{eqnarray*}
where:
\begin{itemize}
\item $idle$ is the phase during which the program runs without using the graphic card
\item $active$ is the phase during which the memory of the GPU is allocated but not used
\item $pause$ is the phase during which the GPU is not used between two computations
\item $end$ is the phase during which the GPU is not used at the end of the execution
\item $n$ is the size of the vectors
\item ${\hat{f}_N^{time,copy}}$ and ${\hat{f}_N^{power,copy}}$ are the unitary transfer computing time and power function
\item ${\hat{f}_N^{time,sum}}$ and ${\hat{f}_N^{power,sum}}$ are the unitary running time and power of the addition of vectors function
\item ${\hat{f}_N^{time,product}}$ and ${\hat{f}_N^{power,product}}$ are the unitary time and power of the product function
\end{itemize}

With these equations, when you develop an algorithm, you are able to have an idea of the energy consumption of this algorithm if you execute it on a specific hardware.

Using our experimental protocol and the data that we measured, we computed these equations for our test bench equipped with an Nvidia GTX275 GPU.
For example for the product, with $n$ being its size:
\[
{\hat{f}_5^{time,product}} = z\mapsto \sum_{k=1}^{5}\frac{\hat{a}_k}{z^k} + \hat{a}_\infty
\]
with
\[
  \left\{
  \begin{array}{cccccccc}
  &\hat{a}_1 &= &0.0000003 &\hat{a}_2 &= &-0.0000027\\
  &\hat{a}_3 &= &0.0000138 &\hat{a}_4 &= &-0.0000362\\
  &\hat{a}_5 &= &0.0000393 &\hat{a}_\infty &= &-9.67\times 10^{-9}\\
  \end{array}
  \right.
\]
and for the sum:
\[
{\hat{f}_5^{power,sum}} = z\mapsto \sum_{k=1}^{5}\frac{\hat{a}_k}{z^k} + \hat{a}_\infty
\]
with
\[
  \left\{
  \begin{array}{ccccccc}
  &\hat{a}_1 &= &0.0000453 &\hat{a}_2 &= &-0.0004421\\
  &\hat{a}_3 &= &0.0021315 &\hat{a}_4 &= &-0.0051157\\
  &\hat{a}_5 &= &0.0049735 &\hat{a}_\infty &= &-0.0000018\\
  \end{array}
  \right.
\]

\subsection{Application to gravity equations}

In this section, we report numerical results where execution times are in seconds, power in Watts, and corresponding energy consumption in Joule. We apply the experimental protocol we developed to linear algebra operations such as addition of vectors, element-wise product or element by element product, dot product and sparse matrix-vector multiplication (SpMV).
SpMV is known to be the most consuming operation in terms of computing time~\cite{cheikahamed:2012:inproceedings-2}~\cite{GPU:BG:2009}. After finding the execution time and energy consumption of these operations, we analyze and evaluate the solution of large size linear systems on GPU with the Conjugate Gradient (CG) algorithm, which is an iterative Krylov method, and is well suited for this type of problem with symmetric positive-definite matrices as demonstrated in~\cite{cheikahamed:2013:inproceedings-3}~\cite{GPU:BFGS:2005}~\cite{GPU:CSVGM:2014}.

\begin{figure}[!ht]
\centering
\scalebox{0.7}{
  \begin{tikzpicture}
    \begin{axis}[
      height=8cm,
      width=10cm,
      xlabel=$size$,
      xmajorgrids, ymajorgrids,
    ]
    \addplot[line width=3pt,color=red,mark=*] table[x index=0,y index=1,col sep=comma] {data/z_spmv.txt};
    \addlegendentry{Alinea}

    \addplot[line width=3pt,color=black,mark=triangle,style=dashed] table[x index=0,y index=4,col sep=comma] {data/z_spmv.txt};
    \addlegendentry{Cusp}
    \end{axis}
  \end{tikzpicture}
  \begin{tikzpicture}
    \begin{axis}[
      height=8cm,
      width=10cm,
      xlabel=$size$,
      xmajorgrids, ymajorgrids,
    ]
    \addplot[line width=3pt,color=red,mark=*] table[x index=0,y index=3,col sep=comma] {data/z_spmv.txt};
    \addlegendentry{Alinea}
    \addplot[line width=3pt,color=black,mark=triangle,style=dashed] table[x index=0,y index=6,col sep=comma] {data/z_spmv.txt};
    \addlegendentry{Cusp}
    \end{axis}
  \end{tikzpicture}
  }
  \caption{Double precision sparse matrix-vector multiplication (CSR format), [top-left: time in seconds (s), top-right: energy  in Joule (J)]}
  \label{fig:tikz:z_spmv}
\end{figure}
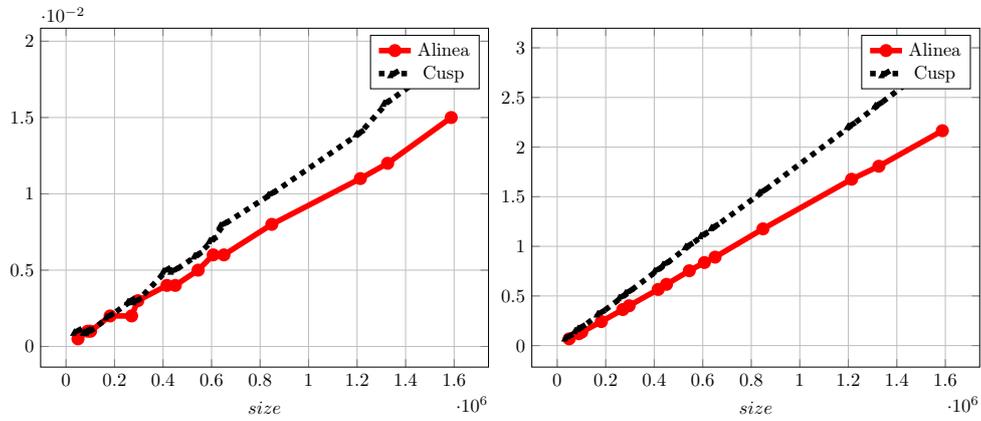

The matrices used in these experiments arises from the finite element discretization of the gravity equations~\cite{ahamed2013stochastic},~\cite{cheikahamed:2013:inproceedings-4}, and are stored in Compressed-Sparse Row, and  The performance of SpMV strongly depends on the structure of non-zero values and also on the memory management as shown in~\cite{ref:cheikahamed:2012:inproceedings-1:DARG:2013}~\cite{ref:cheikahamed:2012:inproceedings-1:DARG:2013:b}. In this experiment, we use \emph{Alinea}, our research group library, which implements most of linear algebra operations on CUDA, using an auto-tuning technique of threading distribution~\cite{cheikahamed:2012:inproceedings-2}. We also compare \emph{Alinea} library with \emph{Cusp} library~\cite{GPU:CUSP:2010} for double precision number arithmetics.
%
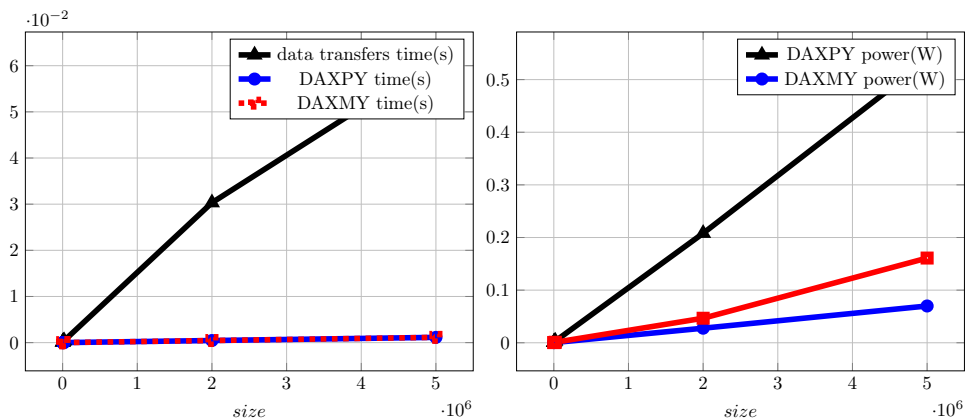
\begin{figure}[!ht]
\centering
\scalebox{0.7}{
  \begin{tikzpicture}
    \begin{axis}[
      height=8cm,
      width=10cm,
      xlabel=$size$,
      xmajorgrids, ymajorgrids,
    ]
    \addplot[line width=3pt,color=black,mark=triangle] table[x index=0,y index=1,col sep=comma] {data/z_data_transfer.txt};
    \addplot[line width=3pt,color=blue,mark=*] table[x index=0,y index=1,col sep=comma] {data/z_add_vector.txt};
    \addplot[line width=3pt,color=red,mark=square,style=dashed] table[x index=0,y index=1,col sep=comma] {data/z_ew_product.txt};
    \addlegendentry{data transfers time(s)}
    \addlegendentry{DAXPY time(s)}
    \addlegendentry{DAXMY time(s)}
    \end{axis}
  \end{tikzpicture}
  \begin{tikzpicture}
    \begin{axis}[
      height=8cm,
      width=10cm,
      xlabel=$size$,
      xmajorgrids, ymajorgrids,
    ]
    \addplot[line width=3pt,color=black,mark=triangle] table[x index=0,y index=2,col sep=comma] {data/z_data_transfer.txt};
    \addplot[line width=3pt,color=blue,mark=*] table[x index=0,y index=2,col sep=comma] {data/z_add_vector.txt};
    \addplot[line width=3pt,color=red,mark=square] table[x index=0,y index=2,col sep=comma] {data/z_ew_product.txt};
    \addlegendentry{DAXPY power(W)}
    \addlegendentry{DAXMY power(W)}
    \end{axis}
  \end{tikzpicture}
}
  \caption{Time (s) and Power (in Watt) for double precision data transfer}
  \label{fig:tikz:basic_operations}
\end{figure}
In Figure~\ref{fig:tikz:basic_operations}, we report respectively the execution time in seconds and the power consumption in Watt (W) for data transferring from CPU to GPU, the addition of vectors (DAXPY) and the element-wise (DAXMY) operations.
Figure~\ref{fig:tikz:z_dot} and Figure~\ref{fig:tikz:z_spmv} respectively give the execution time in seconds and the energy consumption in Joule (J) for the dot product operation and the SpMV operation.
\begin{figure}[!ht]
\centering
\scalebox{0.7}{
  \begin{tikzpicture}
    \begin{axis}[
      height=8cm,
      width=10cm,
      xlabel=$size$,
      xmajorgrids, ymajorgrids,
    ]
    \addplot[line width=3pt,color=red,mark=*] table[x index=0,y index=1,col sep=comma] {data/z_ddot.txt};
    \addlegendentry{Alinea}

    \addplot[line width=3pt,color=black,mark=triangle,style=dashed] table[x index=0,y index=4,col sep=comma] {data/z_ddot.txt};
    \addlegendentry{Cusp}
    \end{axis}
  \end{tikzpicture}
  \begin{tikzpicture}
    \begin{axis}[
      height=8cm,
      width=10cm,
      xlabel=$size$,
      xmajorgrids, ymajorgrids,
    ]
    \addplot[line width=3pt,color=red,mark=*] table[x index=0,y index=3,col sep=comma] {data/z_ddot.txt};
    \addlegendentry{Alinea}

    \addplot[line width=3pt,color=black,mark=triangle,style=dashed] table[x index=0,y index=6,col sep=comma] {data/z_ddot.txt};
    \addlegendentry{Cusp}
    \end{axis}
  \end{tikzpicture}
  }
  \caption{Double precision dot product, [top-left: time in seconds (s), top-right: energy  in Joule (J)]}
  \label{fig:tikz:z_dot}
\end{figure}
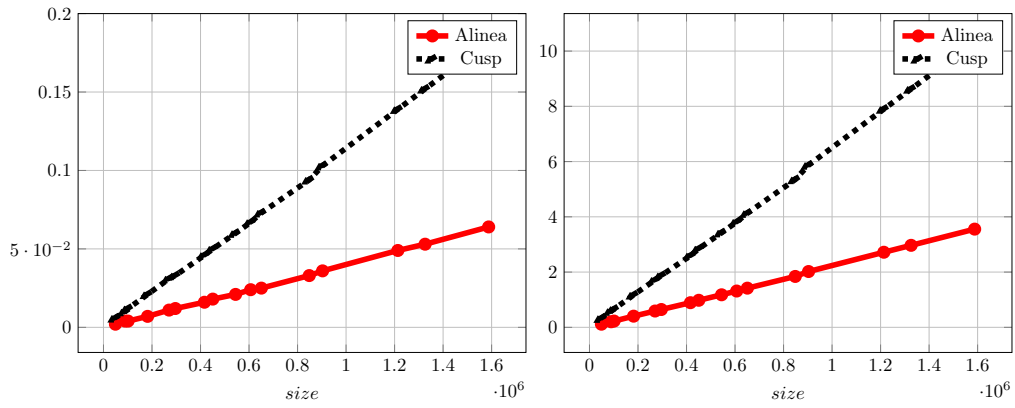

The results of linear algebra operations clearly show the effectiveness of \emph{Alinea} compared to Cusp library for double precision number arithmetics in terms of computing time and energy consumption.

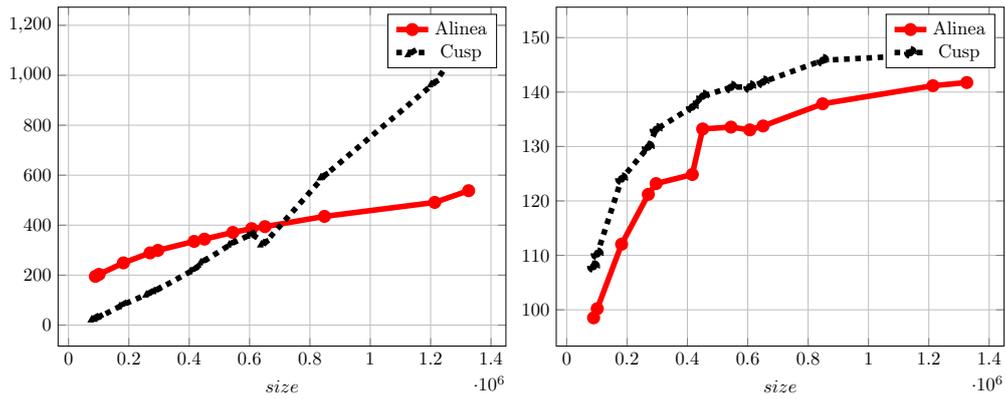
\begin{figure}[!ht]
\centering
\scalebox{0.7}{
  \begin{tikzpicture}
    \begin{axis}[
      height=8cm,
      width=10cm,
      xlabel=$size$,
      xmajorgrids, ymajorgrids,
    ]
    \addplot[line width=3pt,color=red,mark=*] table[x index=0,y index=1,col sep=comma] {data/z_cg.txt};
    \addlegendentry{Alinea}
    \addplot[line width=3pt,color=black,mark=triangle,style=dashed] table[x index=0,y index=4,col sep=comma] {data/z_cg.txt};
    \addlegendentry{Cusp}
    \end{axis}
  \end{tikzpicture}

  \begin{tikzpicture}
    \begin{axis}[
      height=8cm,
      width=10cm,
      xlabel=$size$,
      xmajorgrids, ymajorgrids,
    ]
    \addplot[line width=3pt,color=red,mark=*] table[x index=0,y index=3,col sep=comma] {data/z_cg.txt};
    \addlegendentry{Alinea}

    \addplot[line width=3pt,color=black,mark=*,style=dashed] table[x index=0,y index=6,col sep=comma] {data/z_cg.txt};
    \addlegendentry{Cusp}
    \end{axis}
  \end{tikzpicture}
}

  \caption{Double precision Conjugate Gradient (CSR format), [top-left: time in seconds (s), top-right: energy  in Joule (J)]}
  \label{fig:tikz:z_cg}
\end{figure}

The execution time in seconds and the energy consumption in Joule (J) of the conjugate gradient algorithm are presented in Figure~\ref{fig:tikz:z_cg}. We carry out the CG algorithm with a residual threshold equal to $10^{-6}$.
As we can see in Figure~\ref{fig:tikz:z_cg}, \emph{Alinea} with CSR format is better than Cusp in terms of computing time exept for a range of small size matrices. In terms of energy consumption, \emph{Alinea} is always better than Cusp, even for small matrices.

Previously no preconditionning techniques have been used for the CG algorithm, but additional preconditioners based on domain decomposition methods~\cite{magoules:journal-auth:21,magoules:journal-auth:16} can be used.
For this purpose, special tuned interface conditions between the subdomains are usually defined 
like in the Schwarz method with homogeneous coefficients~\cite{magoules:journal-auth:10,magoules:journal-auth:9} or with heterogeneous coefficients~\cite{magoules:journal-auth:23,magoules:journal-auth:18,magoules:journal-auth:14}.
The associated problem, with the Lagrange multipliers, condensed on the interface between the subdomains, is then solved with an interative algorithm on the CPU.
At each iteration, each subproblem defined in each subdomain is solved with the CG method on the GPU~\cite{ahamed2013stochastic},~\cite{cheikahamed:2013:inproceedings-4}.

\section{Conclusion}

Using graphics card for GPU Computing clearly proved its efficiency of performance in terms of computing time. However, the energy consumed during calculations is far from being negligible.
GPUs are increasingly used in large scale clusters. Because of the high power consumption of these clusters, it is very important to study their energetic aspects, which first of all requires a better understanding of the GPU energetic behavior.
Existing devices measures the energy consumption of the whole computing machine, and not only that of the GPU.

In this paper, we therefore proposed an experimental protocol to measure accurately the power of GPU during computations, and then determine the energy consumed. The presented tests have been performed on a workstation containing two GTX275 GPU cards. We have considered elementary operations, which are the basis of any operations such as classical linear algebra operations, to test and validate our experimental protocol. The evaluation and analysis of the achieved results have enables us to better understand the requirements of the workstation.
Through a good analysis of the energy consumed by a GPU as well as the executed program by determining its number of elementary operations and its execution time, an energy-efficient hardware architecture of the cluster can be consequently identified to find the most efficient HW/SW codesign, paving the path for an optimal task scheduling algorithm.

Finally, we applied the proposed process to solve real problems arising from gravity models using our implementation \emph{Alinea}, and we compared it to Cusp library. The results we obtained outline the robustness, performance and efficiency of our implementation for double precision computation in terms of energy consumption.

\bibliography{bib/hpcc2014_energy,bib/MAGOULES-JOURNAL1,bib/MAGOULES-PROCEEDINGS1}
\bibliographystyle{abbrv}

\end{document}